\documentclass[twocolumn]{aastex631}

\usepackage{amsmath}
\usepackage{wrapfig}

\usepackage{CJK}

\begin{document}
\begin{CJK*}{UTF8}{gbsn}

\title{Dominance of 2-Minute Oscillations near the Alfv\'en Surface}

\author[0000-0001-9570-5975]{Zesen Huang (黄泽森)}
\affiliation{Department of Earth, Planetary, and Space Sciences, University of California, Los Angeles, CA, USA}
\author[0000-0002-2381-3106]{Marco Velli}
\affiliation{Department of Earth, Planetary, and Space Sciences, University of California, Los Angeles, CA, USA}
\author[0000-0002-2582-7085]{Chen Shi (时辰)}
\affiliation{Department of Earth, Planetary, and Space Sciences, University of California, Los Angeles, CA, USA}
\author[0000-0003-3908-1330]{Yingjie Zhu (朱英杰)}
\affiliation{Physikalisch Meteorologische Observatorium, Davos, World Radiation Center (PMOD/WRC), 7260 Davos, Switzerland}
\author[0000-0003-4177-3328]{B. D. G. Chandran}
\affiliation{Space Science Center and Department of Physics, University of New Hampshire, Durham, NH 03824, USA}
\author[0000-0002-4625-3332]{Trevor Bowen}
\affiliation{Space Sciences Laboratory, University of California, Berkeley, CA 94720-7450, USA}
\author[0000-0002-2916-3837]{Victor R\'eville}
\affiliation{IRAP, Universit\'e Toulouse III—Paul Sabatier, CNRS, CNES, Toulouse, France}
\author[0000-0002-9954-4707]{Jia Huang (黄佳)}
\affiliation{Space Sciences Laboratory, University of California, Berkeley, CA 94720-7450, USA}
\author[0000-0001-7205-2449]{Chuanpeng Hou (侯传鹏)}
\affiliation{Institut f\"ur Physik und Astronomie, Universität Potsdam, Potsdam, Germany}
\author[0000-0002-1128-9685]{Nikos Sioulas}
\affiliation{Space Sciences Laboratory, University of California, Berkeley, CA 94720-7450, USA}
\author[0000-0003-2981-0544]{Mingzhe Liu (刘明哲)}
\affiliation{Space Sciences Laboratory, University of California, Berkeley, CA 94720-7450, USA}
\author[0000-0002-1573-7457]{Marc Pulupa}
\affil{Space Sciences Laboratory, University of California, Berkeley, CA 94720-7450, USA}
\author[0000-0002-9951-8787]{Sheng Huang (黄胜)}
\affiliation{Center for Space Physics, Boston University, Boston, MA, USA}
\author[0000-0002-1989-3596]{Stuart D. Bale}
\affiliation{Physics Department, University of California, Berkeley, CA 94720-7300, USA}
\affiliation{Space Sciences Laboratory, University of California, Berkeley, CA 94720-7450, USA}

\begin{abstract}
    Alfv\'en waves, considered one of the primary candidates for heating and accelerating the fast solar wind, are ubiquitous in spacecraft observations, yet their origin remains elusive. In this study, we analyze data from the first 19 encounters of the Parker Solar Probe (PSP) and report dominance of 2-minute oscillations near the Alfv\'en surface. The frequency-rectified trace magnetic power spectral density (PSD) of these oscillations indicates that the fluctuation energy is concentrated around 2 minutes for the ``youngest'' solar wind. Further analysis using wavelet spectrograms reveals that these oscillations primarily consist of outward-propagating, spherically polarized Alfv\'en wave bursts. Through Doppler analysis, we show that the wave frequency observed in the spacecraft frame can be mapped directly to the launch frequency at the base of the corona, where previous studies have identified a distinct peak around 2 minutes ($\sim 8$ mHz) in the spectrum of swaying motions of coronal structures observed by SDO AIA. These findings strongly suggest that the Alfv\'en waves originate from the solar atmosphere. Furthermore, statistical analysis of the PSD deformation beyond the Alfv\'en surface supports the idea of dynamic formation of the otherwise absent $1/f$ range in the solar wind turbulence spectrum.
\end{abstract}


\section{Introduction} \label{sec:intro}

The heliosphere is formed by the supersonic plasma flow known as the solar wind, originating from the Sun. Parker \citep{parker_dynamics_1958} successfully predicted the existence of this supersonic flow, which was later confirmed by \textit{in situ} observations from the Luna 2 \citep{gringauz_study_1962} and Mariner II spacecraft \citep{neugebauer_mariner_1966}. However, it was soon recognized \citep{parker_dynamical_1965} that thermal conduction alone cannot accelerate the solar wind to the observed high speeds ($\gtrsim$ 700 km/s), implying the need for an additional source of momentum and energy that persists into the supersonic regions of the wind \citep{leer_energy_1980}.

Large-amplitude spherically polarized Alfv\'enic fluctuations are ubiquitous in the solar wind \citep{unti_alfven_1968,belcher_large-amplitude_1971,bale_highly_2019,kasper_alfvenic_2019,drake_switchbacks_2021,larosa_switchbacks_2021,shi_patches_2022,telloni_observation_2022,jagarlamudi_occurrence_2023,huang_structure_2023}. As the name says they are characterized by a quasi-constant magnetic magnitude $|B|$, and the tip of the magnetic field vector undergoes a random walk on a sphere \citep{barnes_large-amplitude_1974,tsurutani_nonlinear_1997,matteini_dependence_2014,matteini_alfvenic_2024}. In addition the very high velocity-magnetic field correlation corresponds to that of waves propagating away from the sun. The constant $|B|$ state is an exact solution of the magnetohydrodynamics (MHD) equations, which can propagate without much nonlinear interaction. Thus, the Alfv\'en wave is considered to be a prime candidate for the heating and acceleration of the solar wind \citep{belcher_alfvenic_1971,belcher_stellar_1975,alazraki_solar_1971,hansteen_solar_2012,shi_acceleration_2022,rivera_situ_2024}.

However, the origin of  solar wind Alfv\'enic fluctuations remains a subject of debate \citep{banerjee_magnetohydrodynamic_2021}. Some suggest they are generated by magnetic reconnection in the lower solar corona \citep{bale_solar_2021,drake_switchbacks_2021,bale_interchange_2023}, while others argue that Alfv\'en waves originate deeper in the solar atmosphere \citep{jess_alfven_2009,jess_multiwavelength_2015,morton_alfvenic_2023}. These deeper sources may include the transition region and chromosphere \citep{tian_prevalence_2014,de_pontieu_chromospheric_2007,kuridze_resonant_2008,hou_origin_2024}, or even the photosphere, where they could be generated by p-mode oscillations \citep{ulrich_five-minute_1970} and reach the corona through double mode conversion \citep{cally_alfven_2012,hansen_benchmarking_2012,morton_evidence_2013,morton_investigating_2015,cally_alfven_2017,morton_basal_2019,kuniyoshi_can_2023}, or through buffeting of magnetic flux tubes in ``G band bright points'' by convective motions \citep{cranmer_generation_2005,ballegooijen_heating_2016}. Therefore, leveraging state-of-the-art \textit{in situ} observations is crucial for constraining existing theories.

Parker Solar Probe (PSP) was launched in late 2018 \citep{fox_solar_2016}. Since then, it has been providing \textit{in situ} observations from the magnetically dominated solar corona \citep{kasper_parker_2021,chhiber_alfven_2024}. The unprecedented data from distances to the Sun smaller than $0.25$ AU has ushered in a new era of solar wind turbulence studies (see, e.g., \cite{chen_evolution_2020,shi_alfvenic_2021,zhao_analysis_2021,matthaeus_turbulence_2021,sioulas_magnetic_2022,zank_turbulence_2022,zhang_higher-order_2022,sioulas_evolution_2023,sioulas_magnetic_2023,dunn_effect_2023,larosa_relation_2023,mcintyre_properties_2023}, and \cite{raouafi_parker_2023} for a recent review). In the standard solar wind turbulence model for the fast wind \citep{bruno_solar_2013}, the trace magnetic power spectral density is characterized by a ``$1/f$-inertial'' double power law, where the $1/f$ range at low frequencies is known as the energy-containing range. However, one of the surprising discoveries of PSP is the systematic absence of the low-frequency $1/f$ range near the Alfv\'en surface \citep{huang_new_2023,davis_evolution_2023}, i.e. the region where the solar wind flow overtakes the Alfv\'en speed as it accelerates outwards \citep{kasper_parker_2021,chhiber_alfven_2024}. Instead, the turbulence spectrum is predominantly characterized by a ``shallow-inertial'' double power law, where the shallow part scales like $f^{-0.5}$. Beyond, in the super-Alfv\'enic wind, the $1/f$ range gradually forms between the shallow and inertial part of the spectrum, creating a ``shallow-$1/f$-inertial'' triple power law. As the solar wind travels outwards, the spectrum will eventually transforms into the traditional ``$1/f$-inertial'' scaling. As will be elaborated in this paper, the ``shallow-inertial'' scaling and its deformation have significant implications on the distribution of fluctuation energy in the spectrum.

In this paper, we report the dominance of 2-minute oscillations in the trace magnetic power spectral density near the Alfv\'en surface. The paper is organized as follows: In \S 2, we introduce the data from PSP and examine the implications of the dominant ``shallow-inertial'' double power law near the Alfv\'en surface. We also define a statistical factor, $f_{mid}$, to capture the radial deformation of the spectrum. In \S 3, we establish a connection between the temporal signals measured in the spacecraft frame and the waves launched from the base of the corona. In \S 4, we show the radial evolution of $1/f_{mid}$ in the solar wind. In \S 5, we discuss the implications of our results and conclude our study.

\section{Data and Methodology}\label{sec:data}

\begin{figure*}
    \includegraphics[width=1.0\textwidth]{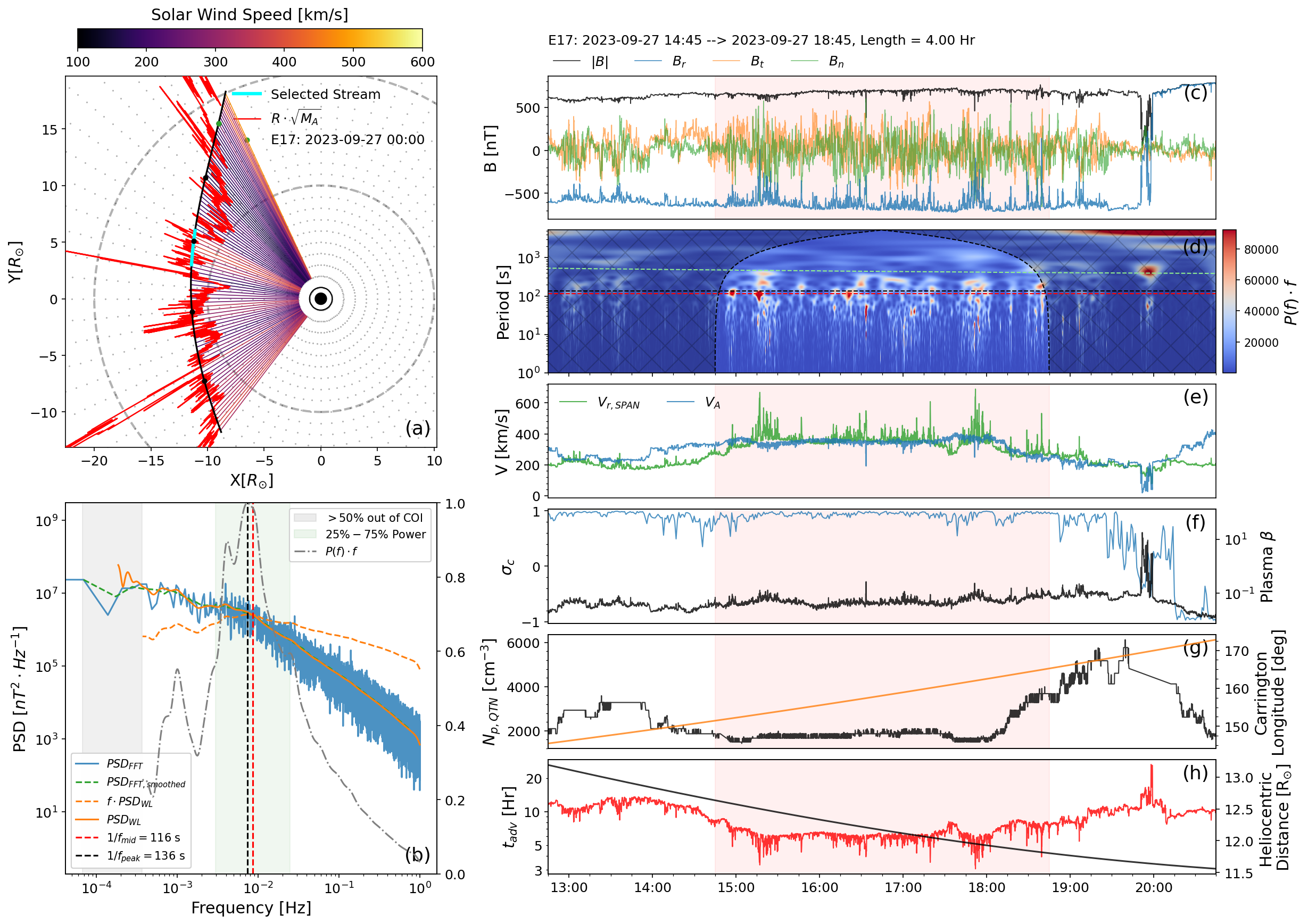}
    \caption{
        (a) Trajectory of PSP in the solar corotating frame.
        (b) Trace magnetic power spectrum density (PSD) of the selected interval. 
        (c) R, T, N components and magnitude of the magnetic field. The selected interval is highlighted with the vertical pink area.
        (d) Trace wavelet scalogram of the magnetic field from the selected interval. 
        (e) Radial solar wind $V_r$ and Alfv\'en speed $V_A$.
        (f) Cross helicity $\sigma_c$ (left, blue) and plasma $\beta=2\mu_0 P/B^2$ (right, black). 
        (g) Proton density $n_p$ (left, black) and Carrington Longitude (right, orange)
        (h) Advection time $\tau_{adv}=(R-R_{\odot})/V_r$ (left, red) and Helio-radial distance $R$ of the spacecraft (right, black). 
    }
    \label{fig:example}
\end{figure*}

\subsection{Data}

We use magnetic field data from the fluxgate magnetometer in the FIELDS instrument suite \citep{bale_fields_2016,bowen_merged_2020} and plasma moments from SPAN-ion in the SWEAP instrument suite \citep{kasper_solar_2016}. In this study, we treat the radial component of the first-order proton moment as the solar wind speed. On PSP, electron number density data from Quasi-Thermal-Noise (QTN) spectroscopy serve as a critical calibration standard for scientific analysis \citep[e.g.,][]{kasper_parker_2021,zhao_analysis_2021,liu_solar_2021,liu_determination_2021,liu_total_2023,zheng_solar_2024}, especially given that the alpha-particle abundance is often negligible (2\% - 6\%) \citep{mostafavi_alphaproton_2022}. Therefore, we use electron number density data derived from the QTN technique \citep[e.g.,][]{moncuquet_first_2020,kruparova_quasi-thermal_2023} as a proxy for proton number density in the solar wind.

\subsection{Example Interval and Methodology}

Figure \ref{fig:example} shows an example interval from PSP's E17 perihelion. The interval of interest is highlighted with the pink shade from panels (c) to (h). In panel (c), we show the RTN components and magnitude of the magnetic field vector. In panel (d), we show the time-periodic trace magnetic spectrogram compiled using wavelet transformation. To convey visually the contributions to the total fluctuation energy from different range of $\ln(f)$, the power spectral density $P(f)$ is rectified with $f$ because $P(f)\ \mathrm{d} f= f P(f)\ {\rm d} \ln f$ \citep[see also][]{liu_rectification_2007}. The black dashed curve marking the boundary of the shaded area is the Cone of Influence (CoI), which indicates the region where the wavelet falls out of the boundary for a given frequency. In panel (e), we show both the radial component of the bulk proton velocity $V_r$ and the local Alfv\'en speed $V_A = |B|/\sqrt{\mu_0 m_p n_p}$ (where $n_p$ is assumed to be the same as $n_e$ estimated from QTN). For this specific interval, $V_r/V_A \simeq 1$, indicating that PSP is situated in the neighborhood of the Alfv\'en surface\citep{kasper_parker_2021,chhiber_alfven_2024}. From panel (f) to (h), we show various plasma parameters, including cross helicity $\sigma_c$ ($\sigma_c = \frac{{z^+}^2 - {z^-}^2}{{z^+}^2 + {z^-}^2}$, where $z^\pm = \delta \vec V \mp \delta \vec B/\sqrt{\mu_0 m_p n_p}$ are the Els\"asser variables), plasma $\beta$ ($\beta = 2\mu_0 P/B^2$), proton number density $n_p$, Carrington longitude, PSP's heliocentric distance $R$, and advection time $t_{adv} = (R - R_{\odot})/V_r$.

Within the interval, the amplitudes of the magnetic vector fluctuations are very large, but the magnetic field magnitude remains relatively constant. Meanwhile, $\sigma_c$ remains very close to unity. The near-perfect anti-correlation between $B_r$ (blue curve, panel (c)) and $V_r$ (green curve, panel (e)) indicates that the fluctuations are unidirectional, outward-propagating, large-amplitude spherically polarized Alfv\'en waves. Recently, these waves have been reported as the major energy source for heating and acceleration of the solar wind \citep{rivera_situ_2024}.

The trajectory of PSP is plotted in the Carrington corotating frame (with the 27-day Carrington rotation subtracted) in panel (a), where the interval of interest is highlighted with the cyan bar (``selected stream''). The local radial solar wind speed ($V_r$ in panel (e)) is shown by the color of the radial lines. To indicate the location of the Alfv\'en surface, the Alfv\'en Mach number ($M_A = V_r/V_A$) is visualized as the red curve. Whenever the red curve dips below the black curve, PSP is inside the magnetically dominant solar corona. To aid the readers, the starts of each day are shown as black dots, the black dashed circles are plotted every 10 $R_{\odot}$, and the radial dotted lines are plotted at 5-degree intervals in Carrington longitude.

Following previous work \citep{huang_new_2023}, in this study we primarily focus on the trace power spectral density ($PSD$) of the magnetic field vector, shown in panel (b). The $PSD_{FFT}$ (blue curve) is estimated using Fast Fourier Transformation (FFT). Due to the mathematical properties of FFT, $PSD_{FFT}$ is naturally noisy. Following the procedure used in previous works \citep{sioulas_evolution_2023,sioulas_magnetic_2023,huang_new_2023}, we smooth $PSD_{FFT}$ with a sliding window with a frequency factor of 2, and the smoothed spectrum is $PSD_{FFT, smoothed}$ (green dashed line). However, the trajectory in panel (a) shows that PSP travels very quickly during perihelion. For the interval of interest (cyan bar), PSP traversed about 14 degrees over 4 hours. Thus, the time series measured by PSP around perihelion is highly non-stationary, violating one of the fundamental assumptions of Fourier transformation. The non-stationary effects primarily influence the low-frequency part of the spectrum. To assess the extent of these effects, we plot the averaged wavelet PSD ($PSD_{WL}$), shown as the orange line (averaged over the data within the CoI in panel (d)). Clearly, $PSD_{WL}$ and $PSD_{FFT, smoothed}$ overlap perfectly in the high-frequency range and begin to diverge around $10^{-3.5}$ Hz, where a significant portion of the data fall outside the CoI. Consequently, we rely on the spectrum down to the frequency where more than 50\% of the data fall outside the CoI ($f_{CoI, 50\%}$), indicated by the gray shade in panel (b). For the remainder of this work, we will use $PSD_{WL}$ as the "primary" PSD of the magnetic field time series (henceforth referred to as $PSD_{WL}$ or $P(f)$ interchangeably), and we will extract factors from $PSD_{WL}$.

$PSD_{WL}$, as mentioned previously, is characterized by a ``shallow-inertial'' double power-law scaling. In the low-frequency range, the spectrum is shallower than $1/f$, and in the high-frequency range, the scaling is consistent with established turbulence phenomenology theories \citep[e.g.,][]{kraichnan_inertialrange_1965,iroshnikov_turbulence_1964,goldreich_toward_1995,goldreich_magnetohydrodynamic_1997,boldyrev_spectrum_2005,boldyrev_spectrum_2006,chandran_intermittency_2015}, and hence it is identified as the turbulence inertial range. One interval with a spectrum shallower than $1/f$ was seen in Helios observations of fast solar wind at 0.3 AU \citep{tu_mhd_1995,chandran_parametric_2018}. Recently, based on the new observations from PSP at much closer distances to the Sun ($R < 0.3$AU), this type of spectrum unexpectedly dominates, and it gradually develops into the typical ``$1/f$-inertial'' spectrum as $R$ increases \citep{huang_new_2023,davis_evolution_2023}.

The ``shallow-inertial'' double power-law has an interesting property. It is well known that $1/f$ scaling indicates that energy is distributed evenly across different frequencies on a logarithmic scale \citep{keshner_1f_1982}. In other words,
\begin{eqnarray}
    \int_{f_1}^{f_2} \frac{1}{f}\ \mathrm{d}f = \ln(f_2/f_1)
\end{eqnarray}
is dependent only on the ratio of the integration bounds, i.e., $f_2/f_1$. Given that the spectrum scales as $P(f) = f^{-\alpha}$, if $\alpha > 1$, the fluctuation energy concentrates at the low-frequency end; and when $\alpha < 1$, the energy concentrates at high frequencies. Consequently, the ``shallow-inertial'' scaling indicates that the energy concentrates around the ``bend''. To substantiate our argument, we consider a fixed window in $\ln(f)$. The energy stored in the moving window is:
\begin{eqnarray}
    \int P(f)\ \mathrm{d} f = \int P(f) f\ \mathrm{d} \ln(f)
\end{eqnarray}
Thus, the rectified spectrum $P(f)\cdot f$ curve is proportional to the "energy density" of frequency $f$ on a logarithmic scale. Additionally, the area under the $P(f)\cdot f$ curve in a linear-log plot represents the relative amount of fluctuation energy for a given frequency range.

In panel (b), $P(f)\cdot f$ is plotted in both logarithmic scale (orange dashed curve) and linear scale (black dotted-dashed curve, right-hand side twin axis). The logarithmic plot clearly shows that the turbulence scaling is shallower than $1/f$ in the low-frequency range ($f < 10^{-2}$ Hz, orange curve goes up) and steeper than $1/f$ in the high-frequency inertial range ($f > 10^{-2}$ Hz, orange curve goes down). Thus, $P(f) \cdot f$ peaks at $1/f_{peak} = 136 \, \text{s}$, indicated by the vertical black dashed line. The area under the black dotted-dashed $P(f)\cdot f$ curve represents the relative amount of total fluctuation energy. A green shade is overlaid to indicate the frequency range containing the middle 50\% of the total fluctuation energy, estimated from the highest frequency $f_0$ down to the right-hand edge of the gray shade $f_{CoI, 50\%}$. In other words, the area under the $P(f)\cdot f$ curve in the green shade contains 50\% of the total area, with the right side up to $f_0$ containing 25\%, and the left side down to $f_{CoI, 50\%}$ containing the remaining 25\%. The middle frequency of the green shade is named $f_{mid}$ and is calculated to be $1/f_{mid} = 116 \, \text{s}$. Both $f_{mid}$ and $f_{peak}$ are indicated by red and black dashed lines in panels (b) and (d).

In this case, the values of $f_{peak}$ and $f_{mid}$ are very close, and the majority of the fluctuation energy concentrates around them (50\% within the green shade). Thus, both can be considered as a ``representative frequency'' of this interval. However, $f_{peak}$ is very sensitive to the specific choice of interval. For example, when the spectrum gradually deforms from ``shallow-inertial'' to ``$1/f$-inertial'', $f_{peak}$ will rapidly migrate to $f_{CoI, 50\%}$ (the lowest trustable frequency, as $P(f)\cdot f$ monotonically decreases as a function of $f$). On the other hand, $f_{mid}$ will slowly migrate to the left as the spectrum deforms, because it represents the center of the frequency range containing the middle 50\% of the total fluctuation energy (see \cite{davis_evolution_2023} for details of the deformation in E10, and \cite{huang_new_2023} for a statistical analysis). Thus, $f_{mid}$ can capture the overall deformation of the spectrum. In this study, we will use $f_{mid}$ as the primary factor for statistical analysis.

\section{Interpretation of Time Series}
\begin{figure*}
    \centering
    \includegraphics[width=.85\textwidth]{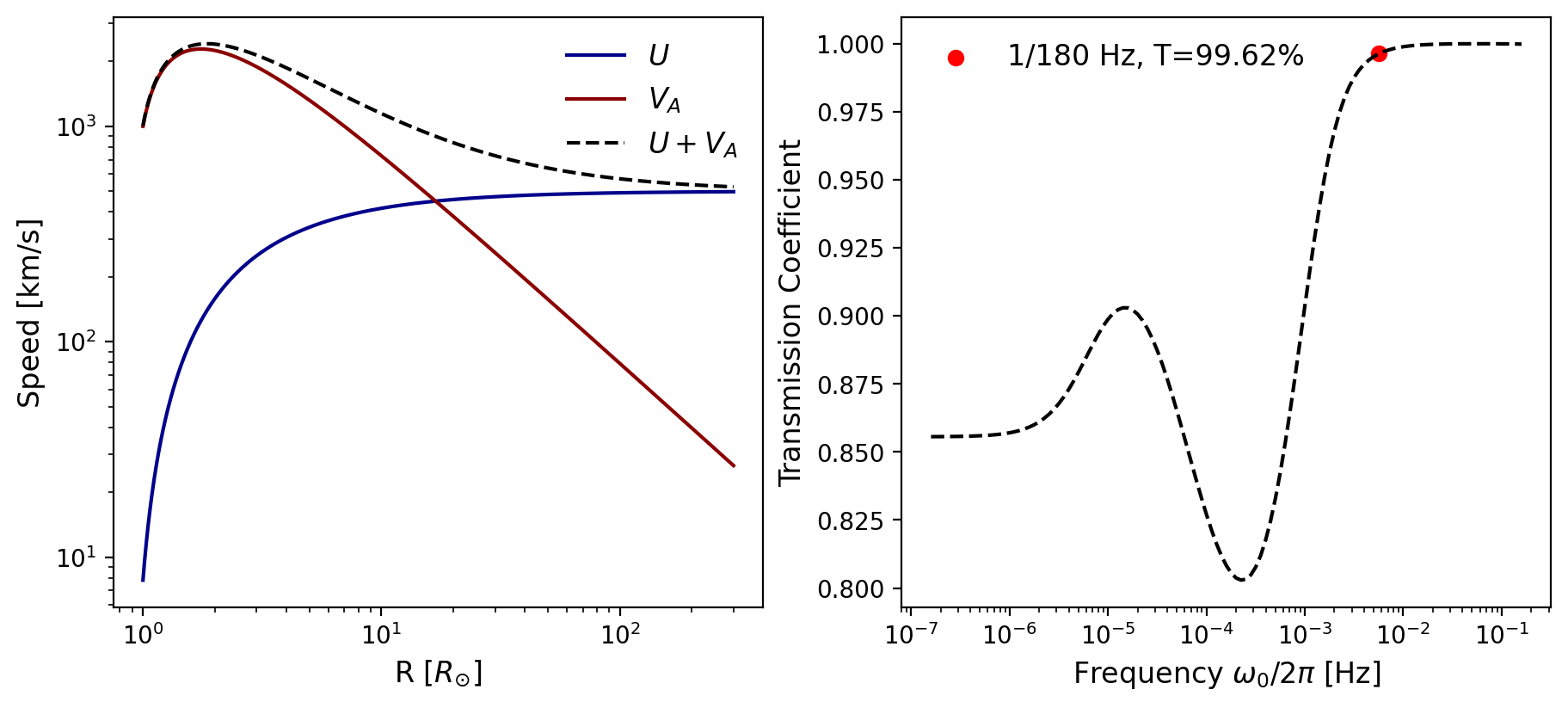}
    \caption{Left: Solar wind and Alfv\'en speed profile. Right: Transmission coefficient of Alfv\'en waves as a function of launch frequency.}
    \label{fig:U_VA_transmission}
\end{figure*}

The frequencies considered in the previous section are measured in the spacecraft frame. Thus, proper interpretation of the spacecraft time series is pivotal to our study. Generally speaking, there are two approaches to interpret the spacecraft time series: (1) directly interpret time series as temporal signals in the spacecraft frame; or (2) invoke the Taylor Hypothesis \citep{taylor_spectrum_1938,perez_applicability_2021} and translate temporal signals into spatial signals. To make meaningful comparisons with anisotropic turbulence phenomenologies \citep[e.g.,][]{kraichnan_inertialrange_1965,iroshnikov_turbulence_1964,goldreich_toward_1995,goldreich_magnetohydrodynamic_1997,boldyrev_spectrum_2005,boldyrev_spectrum_2006,chandran_intermittency_2015}, one must transform the frequency spectrum into the wavevector spectrum. Therefore, the latter approach is more commonly used in the solar wind turbulence community \citep[e.g.,][]{horbury_anisotropy_2012, oughton_anisotropy_2015, duan_anisotropy_2021, adhikari_evolution_2021, zhang_three-dimensional_2022, bandyopadhyay_sub-alfvenic_2022, andres_incompressible_2022, zank_turbulence_2022, sioulas_evolution_2023}.

In this study, we take the first approach. We aim to establish a connection between the waves launched from the base of the corona and the waves measured by PSP near the Alfv\'en surface. In this section, we will show that the Alfv\'en wave frequency observed in the spacecraft frame can be considered identical to the launch frequency for the scenarios considered in this study.

\subsection{One Dimensional Solar Wind Model}

To begin, we consider a simple 1D solar wind model in the solar corotating frame, similar to earlier studies \citep{heinemann_non-wkb_1980,hollweg_alfven_1982,velli_propagation_1993}. In this model, we ignore the perpendicular movement of PSP and consider only its radial speed, $V_{PSP,r}$.

There are three reference frames involved: the Solar surface frame, $K_{\odot}$, the solar wind (intrinsic) plasma frame, $K_{wind}$, which moves at speed $U$ relative to $K_{\odot}$, and the PSP frame, $K_{PSP}$, which moves at speed $V_{PSP,r}$ relative to $K_{\odot}$. Similar to previous 1D models, we launch Alfv\'en waves at the base of the corona with a frequency $f_0$. The frequency of the waves in $K_{wind}$ is:
\begin{eqnarray}
    f_{wind} = f_0 \frac{V_A}{U+V_A}
\end{eqnarray}
where $V_A$ is the local Alfv\'en speed. Similarly, the frequency of the waves in $K_{PSP}$ is:
\begin{eqnarray}
    \begin{aligned}
        f_{PSP} &= f_{wind} \frac{U+V_A-V_{PSP,r}}{V_A}\\
        &= f_0 \frac{U+V_A-V_{PSP,r}}{U+V_A}
    \end{aligned}
\end{eqnarray}
Obviously, $f_{PSP}$ reduces to $f_0$ when $V_{PSP,r} \ll U+V_A$. Due to the spatial gradient of both $U$ and $V_A$ from the base of the corona to PSP, a substantial portion of the Alfv\'en waves could be reflected. In this study, we are particularly interested in the behavior of the Alfv\'en waves observed near the Alfv\'en surface (e.g., Figure \ref{fig:example}). Thus, following \cite{velli_propagation_1993}, we will solve for the transmission coefficient profile for outward-propagating Alfv\'en waves from the base of the corona to the Alfv\'en surface.

Here we use a classical Alfv\'en speed and solar wind speed profile \citep{velli_waves_1991}:
\begin{eqnarray}
    \rho = \rho_0 \frac{\exp\{-\alpha/2\cdot[1-(R_{\odot}/R)]\}}{\{1+\beta[(R/R_{\odot}-1)]\}^2}\\
    V_A = V_{A0} \left(\frac{R_{\odot}}{R}\right)^2\left(\frac{\rho_0}{\rho}\right)^{1/2}\\
    U = \frac{U_{\infty}}{\beta^2} \exp(-\alpha/2)\left(\frac{R}{R_{\odot}}\right)^2\left(\frac{V_A}{V_{A0}}\right)^2
\end{eqnarray}
where $V_{A0}$ is the initial Alfv\'en speed at the base of the corona, $U_{\infty}$ is the asymptotic solar wind speed at infinity, and $\alpha$ and $\beta$ are free parameters. As an example, we choose the following parameters: $\alpha = 0, \beta = 5.6, V_{A0} = 1000 \, \mathrm{km/s}, U_{\infty} = 500 \, \mathrm{km/s}$. Our choice produces an Alfv\'en surface at $12 R_{\odot}$ with $U = V_A = 434 \, \mathrm{km/s}$, consistent with PSP observations in Figure \ref{fig:example}. The profiles of $U$, $V_A$, and $U+V_A$ are shown in Figure \ref{fig:U_VA_transmission}.

To obtain the transmission coefficient profile, we solve the linear propagation equation for the two Els\"asser variables:
\begin{eqnarray}
    \begin{aligned}
    -i \omega z^{\pm} &+ (\vec U \pm \vec V_A) \cdot \nabla z^{\pm} + z^{\mp} \cdot \nabla(\vec U \mp \vec V_A) \\
    &+ \frac{1}{2} (z^{-} - z^{+})\nabla \cdot (\vec V_A \mp \frac{1}{2} \vec U) = 0
    \label{eq:zp and zm}
    \end{aligned}
\end{eqnarray}
where $\omega$ is the angular frequency of the wave in the solar wind frame at distance $R$. The transmission coefficient is defined in terms of the total wave action flux, which is otherwise conserved in the WKB (Wentzel–-Kramers–-Brillouin) limit:
\begin{eqnarray}
    S^\pm = \frac{1}{2} \rho \frac{(z^\pm/2)^2}{\omega_0\frac{V_A}{U\pm V_A}} \cdot (U\pm V_A) \cdot A
    \label{eq:wave action}
\end{eqnarray}
where $A$ is the cross-sectional area of the flux tube. Note that here the wave action density \citep[see e.g.,][]{f_p_bretherton_wavetrains_1968,dewar_interaction_1970} is defined as the energy density divided by the wave frequency $\omega$ in the solar wind frame:
\begin{eqnarray*}
    \frac{1}{2} \rho \frac{(z^+/2)^2}{\omega} = \frac{1}{2} \rho \frac{(z^+/2)^2}{\omega_0 \frac{V_A}{U + V_A}}
\end{eqnarray*}
where $\omega$ is Doppler-shifted downward compared to the launch frequency $\omega_0$ due to the presence of the wind. There are two important properties of (\ref{eq:zp and zm}): First, the net outward-propagating wave action flux is conserved:
\begin{eqnarray}
    S^+ - S^- = S_{\infty} = S_{c}
\end{eqnarray}
where $S_{\infty}$ is the total wave action flux at infinity, and $S_c$ is the flux at the Alfv\'en surface ($S^- = 0$ at the Alfv\'en surface because $U = V_A$). The transmission coefficient for a given launch angular frequency $\omega_0$ is then defined as:
\begin{eqnarray}
    T = \frac{S_{c}}{S^+_0}
\end{eqnarray}
where $S^+_0$ is the total wave action flux at the base of the corona. Second, the singular point at the Alfv\'en surface ($U = V_A$) provides a natural boundary condition for the system. At this singular point, (\ref{eq:zp and zm}) reduces to:
\begin{eqnarray}
    -i \omega z^{-} + \frac{1}{2} (z^{-} - z^{+})\nabla \cdot (\vec V_A + \frac{1}{2} \vec U) = 0
    \label{eq:zm critical}
\end{eqnarray}
Now, combining (\ref{eq:zp and zm}) - (\ref{eq:zm critical}), we integrate the transmission coefficient from the given $U$ and $V_A$ profiles, and the result is shown in the right panel of Figure \ref{fig:U_VA_transmission}.

We may now examine the Doppler shift effects from the radial movements of PSP and the transmission coefficient of the "representative frequency" found in Figure \ref{fig:example}. Around the Alfv\'en surface, $V_{PSP,r} \lesssim 100 \, \mathrm{km/s}$, and $U + V_A \gtrsim 800 \, \mathrm{km/s}$ (Figure \ref{fig:example}(e)). Thus, the radial Doppler shift factor:
\begin{eqnarray*}
    \frac{U + V_A - V_{PSP,r}}{U + V_A} \sim 1
\end{eqnarray*}
is mostly negligible, and the effect is strongest during the fast radial scan phase. The ``representative frequency'' in Figure \ref{fig:example} lies between 2 to 3 minutes. This indicates that the launch frequency of these Alfv\'en waves at the solar surface should be similar. From the right panel of Figure \ref{fig:U_VA_transmission}, we can see that the transmission coefficient for a 3-minute Alfv\'en wave from the base of the corona to the Alfv\'en surface is very close to unity. Thus, based on our 1D solar wind model, the 2-3 minute Alfv\'en waves could be launched from the base of the corona and reach PSP around the Alfv\'en surface without significant reflection.

However, the analysis above has ignored two very important effects: the perpendicular movement of PSP and nonlinear interactions. In the following subsection, we address the first effect, and in the discussion section, we will examine the second effect.

\subsection{Perpendicular Contamination}
\begin{figure*}
    \centering
    \includegraphics[width=1.0\textwidth]{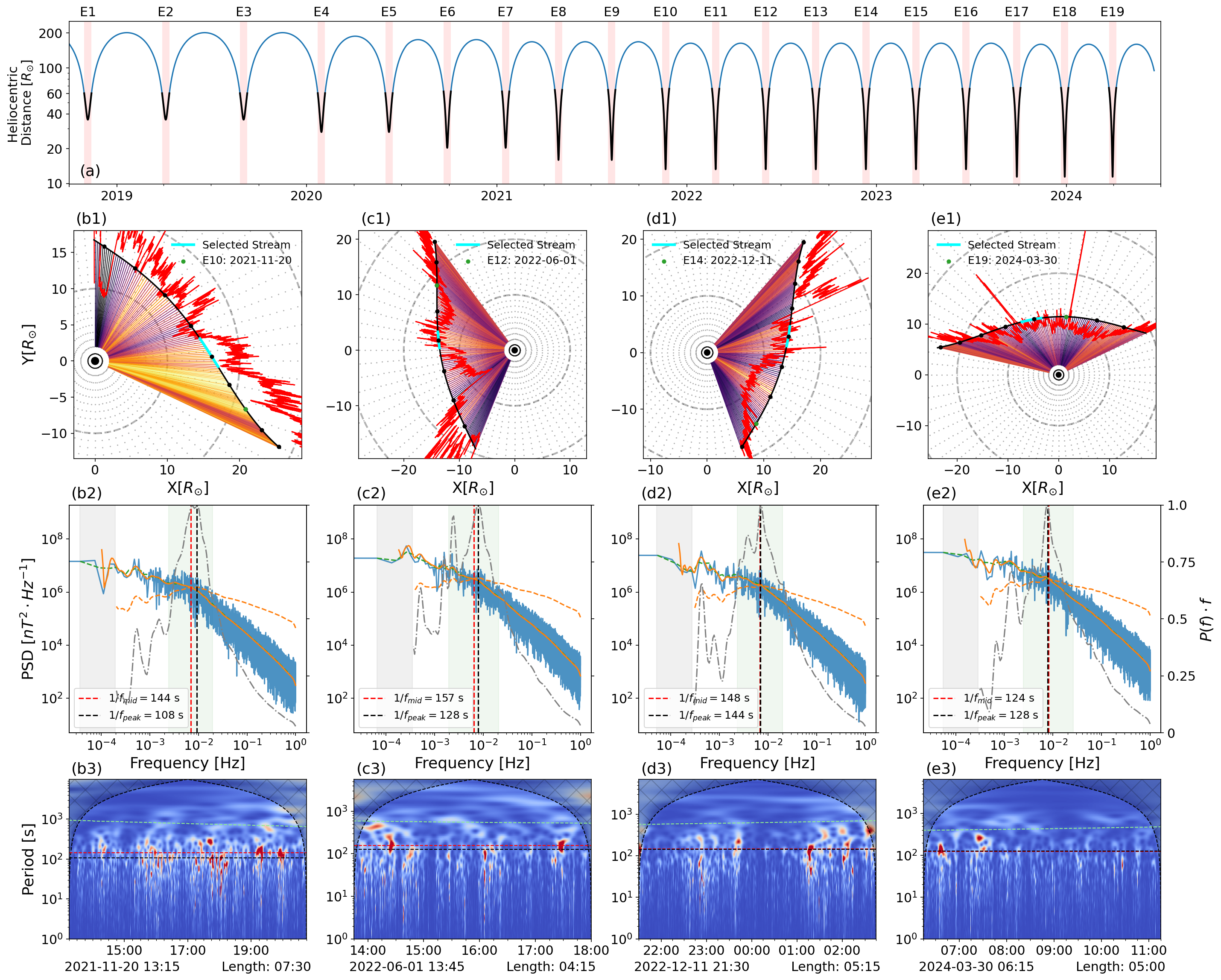}
    \caption{
        (a) Heliocentric distance of PSP as of Jul-01-2024. The time range considered in this study are highlighted with the black lines and pink shaded area.
        (b1-e1) Trajectory of PSP (black line) in Carrington corotating frame. 
        (b2-e2) Power spectrum density of the selected interval.
        (b3-e3) Wavelet scalogram of trace magnetic field power spectrum of the selected interval. 
        The elements in (b), (c) and (d) are identical to Figure \ref{fig:example} (a), (b) and (d).
    }
    \label{fig:combined}
\end{figure*}

The perpendicular motion of the Parker Solar Probe (PSP) causes a Doppler shift that transforms spatial structures into temporal signals, potentially contaminating time series in the spacecraft's frame of reference. This Doppler shift is described by the following relation:
\begin{eqnarray}
    2\pi f = k_\phi \cdot V_{PSP, \phi}
\end{eqnarray}
where $k_{\phi}$ is the wavenumber of the spatial structures in the perpendicular $\phi$ direction. The power spectral density of the perpendicular wavenumber $P(k_\phi)$ is an unknown function. However, it is well established that Alfv\'en waves are guided by the background magnetic field, and around the Alfv\'en surface, this field is predominantly radial. Therefore, the perpendicular spatial signals are likely caused by crossings of intermittent flux tubes. Despite this, it is highly improbable that the concentration of 2-minute oscillations results from perpendicular contamination for two primary reasons.

First, Figure \ref{fig:example}(d) shows that the oscillations are highly localized in the wavelet spectrogram. In other words, the dominant contributors to the peak in the rectified $P(f) \cdot f$ spectrum are the few wave bursts (``red islands'' in panel (d)). This feature of the waves further supports the temporal nature of the signals, rather than contamination from the Doppler shift of spatial structures. To reinforce the universality of this observation, we present four additional example intervals in Figure \ref{fig:combined}. All of these intervals are located near the Alfv\'en surface, and the oscillations remain highly localized in the time-period domain.

Second, for a fixed size at the coronal base, the Doppler-shifted frequency depends on the local ratio $V_{PSP,\phi}/R$. This ratio varies significantly throughout each perihelion since $V_{PSP,\phi}$ increases as PSP approaches perihelion. To illustrate this effect, we calculate an example of the perpendicular contamination frequency based on the size of a solar granule $\lambda_{granule} = 1700$ km \citep{rieutord_power_2010}, assuming a super-radial expansion factor of 3:
\begin{eqnarray}
    f_{c} = \frac{V_{PSP,\phi}}{3\lambda_{granule} \cdot R/R_{\odot}}
    \label{eq:f_photosphere}
\end{eqnarray}
This frequency is represented by green dashed lines in the wavelet spectrograms of Figures \ref{fig:example} and \ref{fig:combined}. In Figure \ref{fig:combined}(b), PSP is nearing the end of its fast radial scan phase, where $R\simeq 17 R_{\odot}$. Here, $V_{PSP,\phi} \simeq 100 \mathrm{km/s}$, yielding $1/f_c \sim 1000\mathrm{s}$. In Figure \ref{fig:combined}(e), as PSP moves closer to perihelion, $R\simeq 12 R_{\odot}$ and $V_{PSP,\phi} \simeq 150\mathrm{km/s}$, leading to $1/f_c \sim 400\mathrm{s}$. Despite these changes, the peaks of the fluctuation energy across all five intervals remain consistent, close to 2 minutes. This suggests that to consistently produce the 2-minute peaks, flux tubes at the coronal base must fine-tune their sizes according to the local ratio of $V_{PSP,\phi}/R$. Additionally, even for locally generated structures, they must adjust their sizes to account for variations in $V_{PSP,\phi}$, which can fluctuate by at least 50\% across the intervals with 2-minute peaks. Achieving such consistent 2-minute peaks under these conditions would be quite remarkable.

In summary, in this section, we have established a connection between the Alfv\'en waves launched from the base of the corona and the Alfv\'en waves measured by PSP near the Alfv\'en surface. The bursty nature of the Alfv\'en waves provides strong evidence that they are temporal signals rather than Doppler-shifted spatial structures. In the following section, we will conduct a statistical analysis of $f_{mid}$ to understand its radial evolution in the solar wind.

\section{Statistical Results}\label{sec:results}

\begin{figure*}
    \centering
    \includegraphics[width=.8\textwidth]{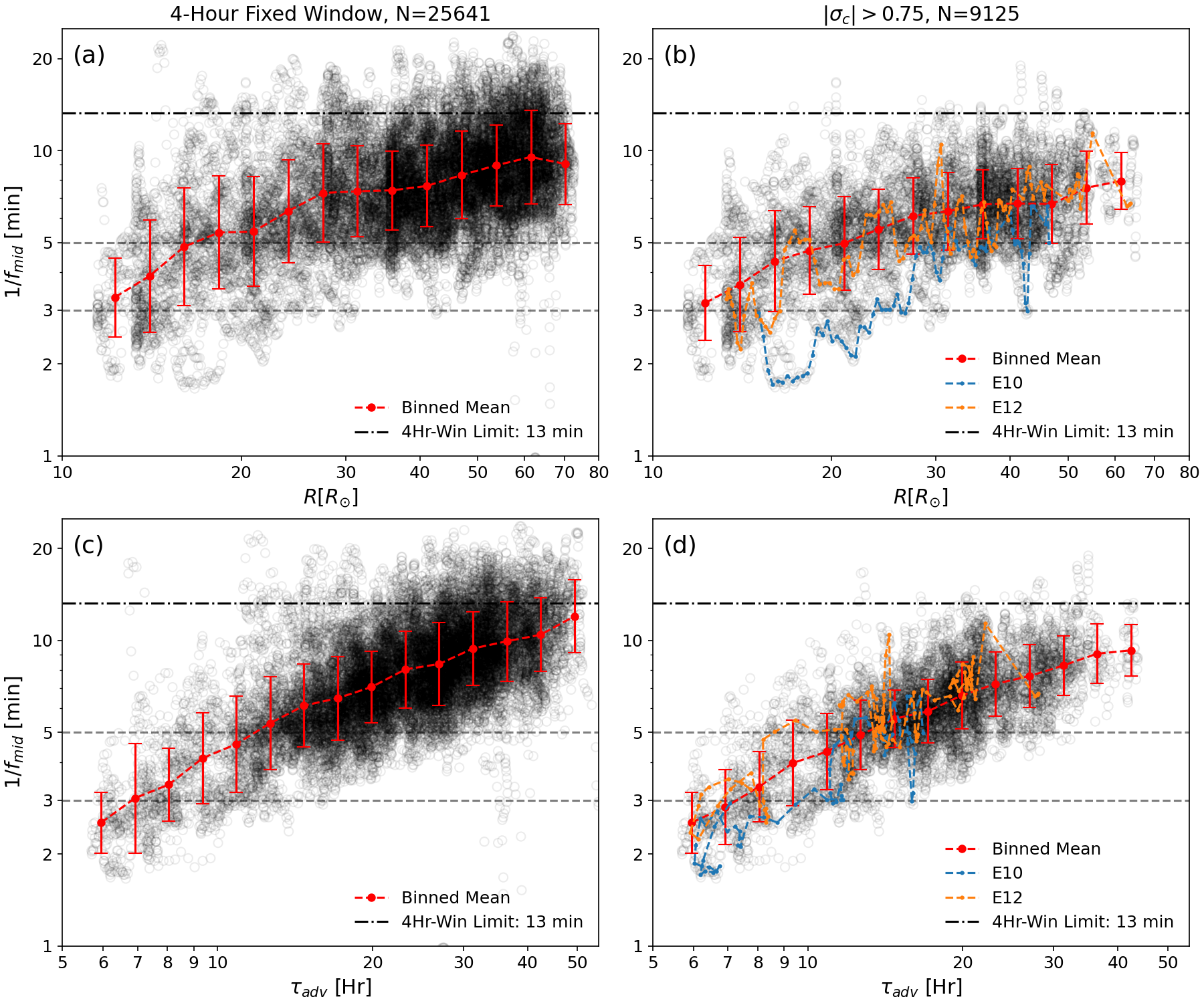}
    \caption{
        (a) Scatter plot of $1/f_{mid}$ as a function of heliocentric distance $R$. The window limit is the $1/f_{mid}$ value estimated from a Kolmogorov $f^{-5/3}$ spectrum. The errorbars indicate one standard deviation in logarithmic scale. (b) Intervals with $|\sigma_c| > 0.75$ are selected. The two known \textit{in situ} coronal holes from E10 and E12 are highlighted. (c) $1/f_{mid}$ is plotted as a function of $t_{adv}=(R-R_{\odot})/V_r$. (d) Intervals with $|\sigma_c| > 0.75$ are selected.
    }
    \label{fig:statistics}
\end{figure*}

To investigate the evolution of $f_{mid}$ in the solar wind, we perform a statistical analysis of data from the first 19 encounters of the Parker Solar Probe (PSP). We consider data from $\pm 7$ days around each perihelion, as highlighted in Figure \ref{fig:combined}(a). A 4-hour sliding window with a fixed 15-minute step is used. For each window, we compute several factors, including $PSD_{WL}$, $f_{mid}$, $f_{peak}$, and other parameters such as $\sigma_c$ and $t_{adv} = (R - R_{\odot})/{V_r}$. In total, 26,623 intervals were selected. As a first step, we cleaned the data, retaining only intervals where $0.005 < \delta t_{adv}/\langle t_{adv} \rangle < 0.2$, leaving us with 25,641 intervals (This criteria is chosen to avoid missing values and abrupt changes in solar wind speed. 982 intervals were filtered out). The values of $1/f_{mid}$ were then sorted by $R$, as shown in Figure \ref{fig:statistics}(a).

The scatter plot in panel (a) appears highly chaotic. This is because, during solar maximum (starting from E10 in late 2021), the ecliptic solar wind consists of a complex mixture of slow and fast wind. The turbulence outer scale in the slow wind is typically observed at very low frequencies $\lesssim 10^{-4} \, \text{Hz}$ \citep{bruno_low-frequency_2019}, whereas in the fast wind, the low-frequency spectral breaks are often found around $10^{-2} \, \text{Hz}$ to $10^{-3} \, \text{Hz}$ \citep{bruno_solar_2013}. As a result, even for the same $R$, the distribution of $1/f_{mid}$ remains highly dispersed. Nevertheless, the binned mean value of $1/f_{mid}$ decreases with decreasing $R$ and drops to about 3 minutes.

A more suitable parameter for sorting $1/f_{mid}$ is the advection time, $t_{adv}$. This parameter can serve as a proxy for the ``age'' of the solar wind plasma, as fast wind naturally takes less time to reach a given $R$ than slow wind. The scatter plot of $1/f_{mid}$ as a function of $t_{adv}$ is shown in panel (c). When sorted by $t_{adv}$, the overall trend becomes much clearer: $1/f_{mid}$ steadily decreases from the saturation value (estimated from a Kolmogorov $f^{-5/3}$ spectrum, see Appendix \ref{sec:window limit} for details) to around 2 to 3 minutes.

What is more intriguing is the evolution of the Alfv\'en waves. In panels (b) and (d), we keep only the intervals where $|\sigma_c| > 0.75$, leaving 9125 intervals. Surprisingly, for both $R$ and $t_{adv}$, the overall trends remain almost unchanged. To probe the evolution of $1/f_{mid}$ within a single stream, we highlight two long intervals from the two confirmed \textit{in situ} coronal holes from the inbounds of E10 \citep{badman_prediction_2023} and E12 \citep{huang_solar_2024} (the dots are connected according to the value of $R$). Following a single stream originating from a coronal hole, $1/f_{mid}$ gradually drops toward 2 minutes as the ``age'' of the plasma decreases.

In summary, the trend of $1/f_{mid}$ as a function of $t_{adv}$ is very clear. For the ``youngest'' solar wind, $1/f_{mid}$ drops to around 2 to 3 minutes. The trend remains largely unchanged when considering only the Alfv\'enic intervals, indicating that this is an intrinsic property of Alfv\'en waves originating from the base of the corona. In the next section, we will provide a detailed discussion on the nonlinear effects and the implications of our results.

\section{Discussion and Conclusions}\label{sec:discussion}

The trend in Figure \ref{fig:statistics}(d) doesn't seem to stabilize with $t_{adv}$. Indeed, both the E10 coronal hole in (d) and the cases in Figure \ref{fig:combined} exhibit $1/f_{mid}$ values smaller than the smallest binned mean from the trend, closer to 2 minutes ($f_{mid} \approx 8.3 \, \mathrm{mHz}$) rather than 3 minutes ($f_{mid} \approx 5.6 \, \mathrm{mHz}$). We argue that the real saturation value of $1/f_{mid}$ is closer to 2 minutes from two perspectives: (1) the actual propagation time of the Alfv\'en waves, and (2) the deformation and evolution of the turbulence spectrum.

The actual propagation time, $t_{prop}$, of the Alfv\'en waves from the base of the corona to the Alfv\'en surface is significantly shorter than $t_{adv}$:
\begin{eqnarray}
    t_{prop} = \int_{R_{\odot}}^{R_c} \frac{dR}{U + V_A}
\end{eqnarray}
where $R_c$ is the heliocentric distance of the Alfv\'en surface. Using the profile in Figure \ref{fig:U_VA_transmission}, we estimate $t_{prop} \approx 1.84 \, \mathrm{hr}$, which is much shorter than the smallest $t_{adv} \approx 5.7 \, \mathrm{hr}$. This indicates that the ``age'' of the Alfv\'en waves in the intervals located at the lower left corner of panel (d) is much ``younger'' than it appears. In contrast, changes in $t_{adv}$ more closely reflect variations in $t_{prop}$, particularly when $t_{adv}$ is large. However, it is clear that a difference of 1.84 hours has little impact on the value of $1/f_{mid}$.

Furthermore, the shift of $1/f_{mid}$ to higher values with increasing $t_{adv}$ likely results from two key effects in solar wind turbulence: (1) the gradual deformation of the spectrum from ``shallow-inertial'' to ``shallow-$1/f$-inertial'', and eventually to ``$1/f$-inertial'' as the solar wind moves away from the Sun \citep{huang_new_2023,davis_evolution_2023}. This transformation, proposed to be caused by the parametric decay-induced inverse cascade of Alfv\'en waves \citep{chandran_parametric_2018}, typically generates a $1/f$ range starting from the ``bend'' that expands toward lower frequencies, gradually shifting $1/f_{mid}$ to the left; and (2) the cascade and dissipation of turbulent fluctuation energy, which consumes energy in the $1/f$ energy-containing range, pushing both the spectral break of the ``$1/f$-inertial'' spectrum and $f_{mid}$ to lower frequencies \citep{tu_mhd_1995, bruno_solar_2013, wu_energy_2020, wu_energy_2021}.

Both mechanisms drive $1/f_{mid}$ to larger values by ``spreading out'' the concentration of fluctuation energy toward lower frequencies. The ``shallow-inertial'' spectrum observed in the most pristine solar wind indicates a concentration of fluctuation energy. To shift $1/f_{mid}$ to even smaller values, the spectrum as a whole would need to move toward higher frequencies, rather than simply deform. To our knowledge, no current turbulence transport theories or models predict such behavior. Therefore, we propose that the true saturation value of $1/f_{mid}$ is likely closer to 2 minutes.

We would also like to emphasize the importance of the rectification of the spectrum: $P(f)\cdot f$. For comparison, \cite{kumar_new_2023} analyzed proton velocity data from the coronal hole outflow in PSP E10 and discovered distinct period peaks around 3, 5, 10, and 20 minutes, consistent with the period peaks in the emission intensity of jetlets at the base of coronal hole plumes and plumelets \citep{uritsky_plumelets_2021,kumar_quasi-periodic_2022}. However, in their wavelet analysis, the PSD was not rectified with the corresponding frequency, as suggested by \cite{liu_rectification_2007}. Due to the high Alfv\'enicity of the coronal outflow, if rectified, the PSD from the proton velocity should resemble the trace magnetic PSD compiled in this study.

Our findings show a striking consistency with the remote sensing spectrum derived from the swaying motions of coronal structures observed by SDO/AIA within coronal holes. Specifically, in Figure 6(c) of \cite{morton_investigating_2015}, the peak of the histogram for Alfv\'en(ic) wave periods falls between 100s and 200s. Furthermore, in Figure 1(d) of \cite{morton_basal_2019}, after rectifying the SDO/AIA coronal hole power spectral density (PSD) with $f$, the true peak of the spectrum emerges clearly at around 8 mHz (or approximately 2 minutes). However, this peak is absent in the PSD compiled from CoMP Doppler velocity time series. The authors noted that due to instrument uncertainties, they were unable to confirm the authenticity of this peak.

Finally, our results establish a strong connection between the Alfv\'en waves measured by PSP near the Alfv\'en surface and those launched from the base of the corona. The dominance of 2-minute Alfv\'en waves near the Alfv\'en surface suggests that the Alfv\'en waves in the solar wind could be the result of leakage from p-mode oscillations, as proposed by previous studies \citep{cally_alfven_2012,hansteen_solar_2012,morton_evidence_2013,morton_investigating_2015,cally_alfven_2017,morton_basal_2019}. However, current evidence \citep{morton_basal_2019} is not yet robust enough to confirm this hypothesis. It is also possible that Alfv\'en waves at the corona's base are generated by convective motions at the photosphere, which shake the footprints of open magnetic field lines \citep{cranmer_generation_2005,ballegooijen_heating_2016}. A detailed discussion to differentiate between these models is, unfortunately, beyond the scope of this study.

Future studies on this matter can focus on two avenues: First, PSP will reach its deepest perihelion in December 2024. The new data from deeper below the Alfv\'en surface will provide valuable insights into the evolution of the turbulence spectrum. Thus, a dedicated cross-validation study between \textit{in situ} observations and remote sensing in the future is of great interest. Second, the origin of the energy-containing $1/f$ range has been debated for decades in the field \citep{keshner_1f_1982,montroll_1f_1982,bak_self-organized_1987,velli_turbulent_1989,matthaeus_low-frequency_1986,matthaeus_density_2007,bemporad_low-frequency_2008,dmitruk_emergence_2011,verdini_origin_2012,matteini_1_2018,chandran_parametric_2018,magyar_phase_2022,meyrand_reflection-driven_2023}. A more detailed study on the evolution of the spectrum in the young solar wind will likely shed new light on this topic.

\section*{Acknowledgements}
The authors acknowledge the following open-source packages: \cite{lam_numba_2015,harris_array_2020,hunter_matplotlib_2007,angelopoulos_space_2019,virtanen_scipy_2020}. Z.H. thanks Pankaj Kumar for valuable discussions. This research was funded in part by the FIELDS experiment on the Parker Solar Probe spacecraft, designed and developed under NASA contract UCB \#00010350/NASA NNN06AA01C, and the NASA Parker Solar Probe Observatory Scientist grant NASA NNX15AF34G, and NASA HTMS 80NSSC20K1275. M.V. acknowledges support from ISSI via the J. Geiss fellowship. B.C. acknowledges support from NASA grant 80NSSC24K0171. C.S. acknowledges supported from NSF SHINE \#2229566 and NASA ECIP \#80NSSC23K1064.

\clearpage
\appendix

\section{Power Spectrum Density and Window Limit}\label{sec:window limit}
\begin{figure*}[h]
    \centering
    \includegraphics[width=0.50\textwidth]{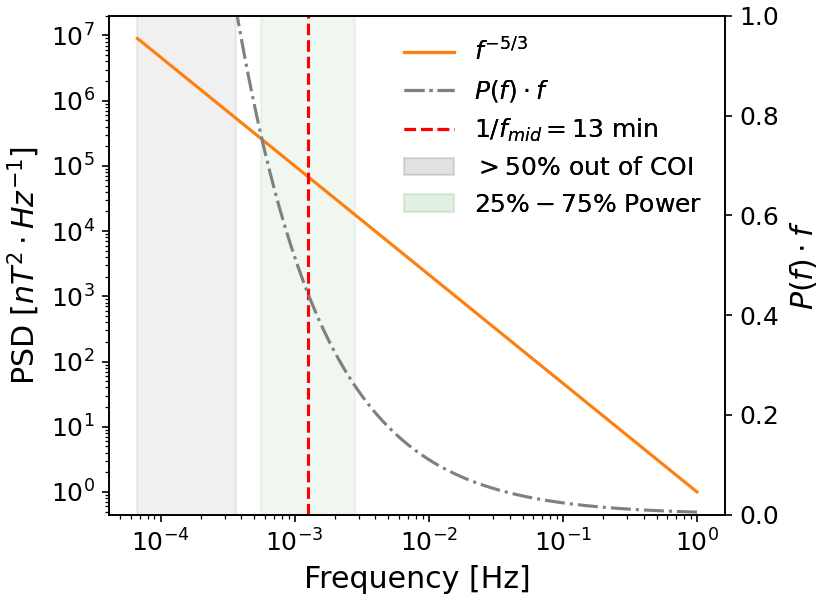}
    \caption{
        Illustration of the saturation $1/f_{mid}$ with a 4-hour window. Orange: Kolmogorov $P(f) = f^{-5/3}$ spectrum; Dotted dashed: Compensated spectrum $P(f)\cdot f$ in linear scale on twin axis; Gray area: Frequency range where $>$50\% of points are out of the Cone of Influence; Green area: frequency range that contains 50\% of fluctuation energy (from 25\% to 75\%).
    }
    \label{fig:window limit} 
\end{figure*}

Figure \ref{fig:window limit} is an illustration of the window limit of 13 minute for the 4-hour window (the dotted dashed line in Figure \ref{fig:statistics}) and the definition of the lines are consistent with Figure \ref{fig:example}(b). An artificial Kolmogorov spectrum $P(f) = f^{-5/3}$ is plotted with the orange line. The middle frequency $f_{mid}$ is defined as the median value of the frequency range which contains the middle 50\% of the total fluctuation energy.

\clearpage
\bibliography{sample631}{}
\bibliographystyle{aasjournal}

\end{CJK*}
\end{document}